\documentclass[twocolumn,showpacs,aps]{revtex4}
\usepackage[dvips]{graphicx}
\usepackage{bm}
\usepackage{mathrsfs}

\begin{document}
\date{\today}

\title{Diffusion in a Curved Tube}
\author{Naohisa Ogawa
\footnote{ogawanao@hit.ac.jp}}
\affiliation{Hokkaido Institute of Technology, Sapporo 006-8585, Japan}
\begin{abstract}
The diffusion of particles in confining walls forming a tube is discussed.
Such a transport phenomenon is observed in biological cells and porous media.
We consider the case in which the tube is winding with curvature and torsion, and the thickness of the tube is sufficiently small compared with its curvature radius. 
We discuss how geomerical quantities appear in a quasi-one-dimensional diffusion equation.
\end{abstract}
\maketitle

\section{Introduction}
To control the transportation of micro- and nanoparticles artificially, it is very important to understand the diffusion properties under the confining walls. These phenomena are encountered in biological cells 
\cite{biological cell} and zeolite \cite{zeolite}, and in catalytic reactions in porous media \cite{porous media}.
For this purpose,  the diffusion properties in confined geometries are discussed by several authors. The diffusion in a membrane with a certain thickness is discussed by Gov \cite{Nir}, Gambin et al. \cite{Gambin},  and Ogawa \cite{ogawa}. The diffusion in general curved manifold is discussed by Castro-Villarreal \cite{Pavel}.  The diffusion in a tube with a varying cross section along the axis (channel model)  is discussed by  Jacobs \cite{Jacob}, Yanagida \cite{yanagida},  Zwanzig \cite{Zwanzig}, Reguera and Rubi \cite{Reguera}, and Kalinary and  Percus \cite{Kalinary}, and as a review, see Burada et al. \cite{review}.  Surprisingly, this channel model is related to the reaction rate theory due to the Smoluchowski equation \cite{review}, \cite{Haenggi}.

In this paper, we discuss the case in which a tube has a fixed cross section but is winding with geometrical properties, namely, curvature and torsion. 
Then, we show that the diffusion in such a tube with a Neumann boundary condition can be expressed by a quasi-one-dimensional diffusion equation with an effective diffusion coefficient that depends on curvature.
This is carried out by integrating a three-dimensional diffusion equation in the cross section of the tube. 
The coefficient depends on the curvature of the central line of the tube.
The physical interpretation of  its curvature dependence is given by analogy to Ohm's law.

By using the obtained equation, we show the mean square displacement (MSD) of torus and helix tubes where the curvature is constant.  
When the curvature depends on position, we show the short time expansion for MSD.

In section 2, we introduce the curvilinear coordinates and related metrics in a winding tube. 
This is carried out by using Frenet-Seret (FS) equations explained in appendix 1.
In section 3, we define the quasi-one-dimensional diffusion field.  
In section 4, the diffusion equation is obtained by using a local equilibrium condition.
This condition is an assumption that the diffusion in the same cross section is completed in a short time, 
which is much smaller than our observed time scale; thus, we may assume that the density is flat on the same cross section. In section 5, we discuss the diffusion equation beyond the local equilibrium condition. In section 6, we calculate MSD from the quasi-one-dimensional diffusion equation and show first two terms in the short time expansion by using curvature and its derivatives. In section 7, the conclusion is given.

\begin{figure}
\centerline{\includegraphics[width=4cm]{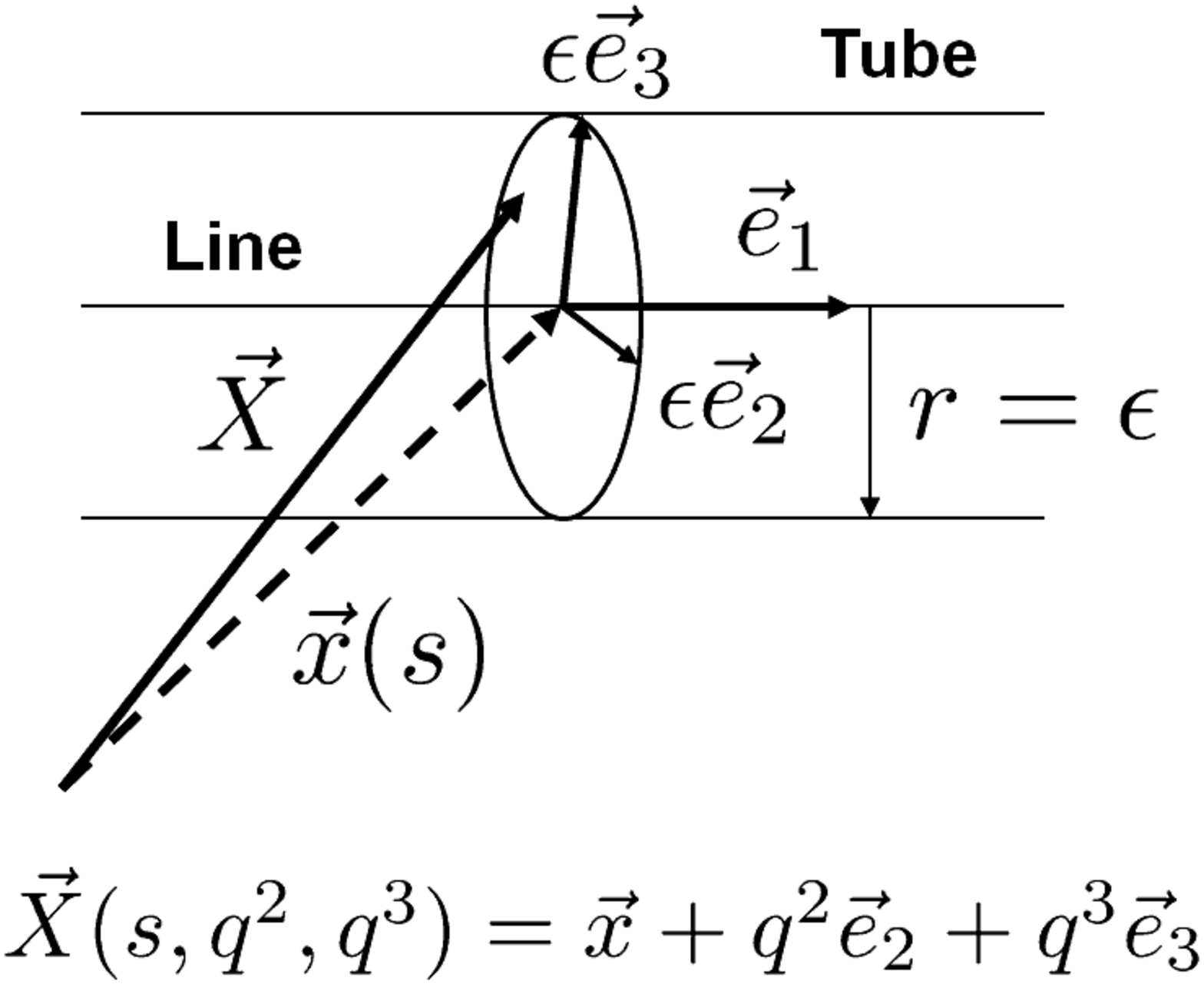}}
\caption{Local orthogonal coordinates $\{s, q^2,q^3\}$ in tube.}
\end{figure}

\section{Metric in Tube}

We consider the quasi-one-dimensional diffusion equation as the limitation process from a three-dimensional usual diffusion equation in a thin tube with a circular cross section.
We set the curved tube with the radius $\epsilon$ in the three-dimensional Euclidean space $R_3$.
The curvilinear coordinates that specify the points in the tube and their bases we use hereafter are as  follows (see figure 1).

$\vec{X}$ is the Cartesian coordinate in $R_3$.
$s~(=q^1)$ is the length parameter along the center line of the tube and $\vec{e}_1$ is its tangential vector. 
$\vec{x}(s)$ is the Cartesian coordinate that specifies the points on the center line.
$q^i$ is the coordinate in the transversal direction $\vec{e}_i$, the small Latin indices $i,j,k,\cdots$ run from 2 to 3 and the Greek indices $\mu, \nu,\cdots$ run from 1 to 3.
We sometimes use the notations  $s = q^1, v = q^2= r \cos \theta,$ and $w = q^3 = r \sin \theta$ to obtain simpler expressions, and we define the area element of the cross section $d \sigma = dv dw = r dr d\theta$.
$\vec{e}_1, ~ \vec{e}_2,$ and $\vec{e}_3$ are the unit basis vectors introduced by the Frenet - Seret equations \cite{FS1}, \cite{FS2} explained in appendix 1.
Then, we identify any points in the tube using

\begin{equation}
\vec{X}(s,q^2,q^3) = \vec{x}(s) + q^i \vec{e}_i(s),
\end{equation}
where $ 0 \leq \mid \vec{q} \mid \leq \epsilon $ with $\mid \vec{q} \mid = \sqrt{(q^2)^2 + (q^3)^2}$.

From this relation, we obtain the curvilinear coordinate system in the tube ($\subset R_3$)
using the coordinate $q^{\mu}=(q^1,q^2,q^3)$ and metric $G_{\mu\nu}$.

\begin{equation}
G_{\mu\nu}=\frac{\partial \vec{X}}{\partial q^{\mu}} \cdot \frac{\partial \vec{X}}{\partial q^{\nu}}.
\end{equation}

$G_{\mu\nu}$ is calculated by using the Frenet - Seret equations (appendix 1).
\begin{eqnarray}
&&G_{\mu\nu} = \nonumber\\
&&\left(
\begin{array}{ccc}
1-2 \kappa v +(\kappa^2 + \tau^2)v^2 +\tau^2 w^2 & -\tau w & \tau v\\
- \tau w & 1& 0\\
\tau v & 0 & 1
\end{array}
\right),\label{eq:metric}
\end{eqnarray}
where $\kappa$ is the curvature and $\tau$ is the torsion defined in appendix 1.
We have nonzero off-diagonal elements due to the existence of torsion.
The determinant of the metric tensor does not depend on torsion.
\begin{equation}
G \equiv \det (G_{\mu\nu}) = (1-\kappa v)^2.
\end{equation}

The inverse metric is given as
\begin{eqnarray}
&&G^{\mu\nu} = \frac{1}{(1-\kappa v)^2} \times  \nonumber\\
&&\left(
\begin{array}{ccc}
1 & \tau w &-\tau v\\
\tau w &  (1-\kappa v)^2 + (\tau w)^2 &-\tau^2 vw\\
-\tau v & -\tau^2 vw & (1-\kappa v)^2 + (\tau v)^2
\end{array}
\right).~
\end{eqnarray}

\section{Diffusion field in Tube}

Let us define a three- dimensional diffusion field by $\phi^{(3)}$ and a three- dimensional Laplace-Beltrami operator with metric tensor (\ref{eq:metric}) by $\hat{\Delta}$.
Then, we obtain the diffusion equation with a normalization condition:
\begin{eqnarray}
&& \frac{\partial \phi^{(3)}}{\partial t} = D \hat{\Delta} \phi^{(3)},\label{eq:diff}\\
&& N = \int \phi^{(3)}(q^1,q^2,q^3)  \sqrt{G}~ d^3 q,\label{eq:norm1}
\end{eqnarray}
where $D$ is the diffusion constant, $G \equiv \det(G_{\mu\nu})$, and $N$ is the number of particles.
Our aim is to construct the effective one-dimensional diffusion equation from the 3D equation above in a small radius limit.

\begin{eqnarray}
&& \frac{\partial \phi^{(1)}}{\partial t} = D \hat{\Delta}^{(eff)} \phi^{(1)},\\
&& N = \int \phi^{(1)}(s) d s,\label{eq:norm2}
\end{eqnarray}
where $\phi^{(1)}$ is the one-dimensional diffusion field and $\hat{\Delta}^{(eff)}$ is the unknown effective 1D diffusion operator that might not be equal to the simple 1D Laplace-Beltrami operator $\partial^2/\partial s^2$.

From two normalization conditions, namely,  (\ref{eq:norm1}) and (\ref{eq:norm2}), we obtain
\begin{eqnarray}
N &=& \int \phi^{(3)}(q^1,q^2,q^3)  \sqrt{G}~ d^3 q,\nonumber \\
&=& \int [\int d q^2 d q^3  (\phi^{(3)} \sqrt{G})] ~ ds, \nonumber\\
&=& \int \phi^{(1)}(s) ~ ds.\nonumber
\end{eqnarray}
The particle number between $s$ and $s+ds$ should be equal in the two fields.
Thus, we obtain
\begin{equation}
\phi^{(1)}(s) = \int \phi^{(3)} \sqrt{G} ~d q^2 d q^3. \label{eq:relation}
\end{equation}

We multiply $\sqrt{G}$ by equation (\ref{eq:diff}) and integrate it by $q^2$ and $q^3$ to obtain
\begin{equation}
\frac{\partial \phi^{(1)}}{\partial t} = D \int (\sqrt{G} \hat{\Delta}) \phi^{(3)} dq^2 dq^3. \label{eq:laplace}
\end{equation}

From the form of the Laplace-Beltrami operator
$$\Delta = G^{-1/2} \frac{\partial}{\partial q^\mu} G^{1/2} G^{\mu\nu} \frac{\partial}{\partial q^\nu},$$

our diffusion equation has the form

\begin{eqnarray}
\frac{\partial \phi^{(1)}}{\partial t} &=& D  \int  \frac{\partial}{\partial q^\mu} G^{1/2} G^{\mu \nu} 
 \frac{\partial}{\partial q^\nu}  \phi^{(3)}  d\sigma \nonumber\\
&=&  D\frac{\partial}{\partial s} \int \frac{1}{\sqrt{G}} (\frac{\partial}{\partial s} - \tau  \frac{\partial}{\partial \theta} ) \phi^{(3)} d\sigma, \label{eq:new}
\end{eqnarray}
where the Neumann boundary condition is used at the second equality.
The torsion appears only when the axial symmetry of $\phi^{(3)}$ is broken as is expected from the definition.

\section{Local Equilibrium Condition}

Now, we suppose the ``local equilibrium condition'' as

\begin{equation}
\frac{\partial \phi^{(3)}}{\partial q^i} =0, ~~i=2,3. \label{eq:lec}
\end{equation}

In the directions $\vec{e}_2$ and $\vec{e}_3$, the local equilibrium holds for a short time $\delta t \sim \epsilon^2/D$. 
When our observation is given in the time scale $t$ satisfying $t ~>> \delta t$, we can assume local equilibrium condition (\ref{eq:lec}). Note that this condition includes the Neumann condition at the boundary of the tube.

From equations (\ref{eq:relation}) and (\ref{eq:lec}), we also obtain
\begin{equation}
\phi^{(3)}= \frac{\phi^{(1)}}{\sigma} , ~~ \sigma \equiv \int \sqrt{G} d\sigma = \pi \epsilon^2. \label{eq:cond}
\end{equation}

Note that  $\phi^{(3)}= \phi^{(1)}/(\sigma \sqrt{G})$ also satisfies condition (\ref{eq:relation}); however, this relation does not satisfy the Neumann condition at the boundary of the tube. 
Then, we obtain

\begin{equation}
\frac{\partial \phi^{(1)}}{\partial t} =    \frac{\partial}{\partial s} D_{eff} \frac{\partial}{\partial s}  \phi^{(1)}, \label{eq:result}
\end{equation}

where

\begin{eqnarray}
D_{eff} &\equiv& \frac{D}{\pi \epsilon^2}  \int \frac{d \sigma}{G^{1/2}}  \nonumber\\
&=& \frac{D}{\pi \epsilon^2} \int \frac{ d\sigma }{1-\kappa q^2} \nonumber \\
&=& 2D \frac{1-\sqrt{1-(\kappa \epsilon)^2}}{ (\kappa \epsilon)^2}\nonumber\\
&=& D \{ 1 + (\frac{\kappa \epsilon}{2})^2 + {\cal O} (\epsilon^4) \}.
\end{eqnarray}

The static solution has the form

\begin{equation}
\phi^{(1)} = C_1 + C_2 \int_0^s \frac{ds'}{D_{eff}(s')}.
\end{equation}

Note that equation (\ref{eq:result}) is consistent with normalization condition (\ref{eq:norm2}).

We also obtain the relation
\begin{equation}
D_{eff} = D <\frac{1}{1-\kappa q^2}>, \label{eq:solution}
\end{equation}
where
$$ < \cdots > = \frac{1}{\pi \epsilon^2}\int d\sigma \cdots.$$

Then, we find a simple interpretation of this relation.
Let us consider the point P on the tube where the curvature is $\kappa$.
We choose two sections near P and discuss the length connecting these two sections (see figure 2). 
 At the coordinate $q^2$, the length between the two sections is given by $s$, which is different from the length of the center line, $\bar{s}$, with the ratio
\begin{equation}
\frac{\bar{s}}{s} =  \frac{1}{1-\kappa q^2}. \label{eq:ratio}
\end{equation}

Next, we should note that the relation between diffusion flow 
and density difference is similar to Ohm's law.
$$ J \sigma = \frac{D \sigma}{s} \Delta N ~ \sim ~ I = \frac{1}{R} V,$$
where $J$ is the flow density, $\sigma$ is the cross section, $D$ is the diffusion constant, $s$ is the distance, and $\Delta N$ is the density difference. In the comparison with Ohm's law, electric current corresponds to $J \sigma$, voltage corresponds to $\Delta N$, and electric conductivity $1/R$ corresponds to $D \sigma /s $. By using the above correspondence, we consider our tube as a bundle of thin tubes (see figure 3).  Thus, it can be seen as a parallel connection of many resistances. The total conductivity is calculated as

\begin{figure}
\centerline{\includegraphics[width=4cm]{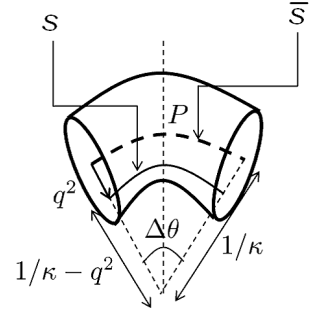}}
\caption{Bent point $P$ of tube: $q^2$ shows distance to the direction of center of curvature.
The center line has the curvature radius $1/\kappa$ and the line with $q^2 \neq 0$ has the curvature radius $1/\kappa - q^2$.}
\end{figure}

\begin{equation}
\frac{1}{R} = \sum_{i=1}^N \frac{1}{R_i} = D \sum_{i=1}^N \frac{\Delta \sigma_i}{s_i} \equiv D_{eff} \frac{\sigma}{\bar{s}}, \label{eq:eff_diff}
\end{equation}
where $ \sigma = \sum_{i} \Delta \sigma_i $.
The last equality shows the definition of the effective diffusion coefficient.
Thus, we obtain
\begin{eqnarray}
D_{eff} &=& \frac{D}{\sigma} \sum_{j=1}^N \frac{\bar{s} }{s_j}\Delta \sigma_j = \frac{D}{\pi \epsilon^2} \int \frac{d \sigma}{1-\kappa q^2} \nonumber\\
&=& D <\frac{1}{1-\kappa q^2}>,
\end{eqnarray}
where (\ref{eq:ratio}) is utilized at the second equality.

\begin{figure}
\centerline{\includegraphics[width=6cm]{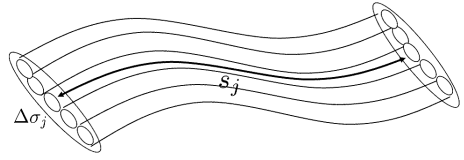}}
\caption{Bundle of tubes: our tube is considered as a bundle of infinitesimally thin tubes with length $s_j$ and cross-sectional area $\Delta \sigma_j$.}
\end{figure}

When the cross section is not circular but quadrangular as seen in figure 4, we obtain 
\begin{eqnarray}
 D_{eff} &=& \frac{D}{W \epsilon} \int^{\epsilon/2}_{-\epsilon/2} dq^2 \int^W_0 dq^1 ~ \frac{1}{1- \kappa q^2}\nonumber\\
&=& \frac{D}{\kappa \epsilon} \ln \mid \frac{1+\kappa \epsilon/2}{1-\kappa \epsilon/2}\mid \nonumber\\
&=& D (1 + \epsilon^2 \kappa^2 /12 + \cdots).
\end{eqnarray}

Note that $D_{eff}$ is an even function of $\kappa$.
This result is consistent with the surface diffusion with the thickness $\epsilon$, 
where one tangential direction is flat and the other has the curvature $\kappa$, similarly to the case of a surface on an elliptic cylinder, which is shown in \cite{ogawa}. 

\begin{figure}
\centerline{\includegraphics[width=4cm]{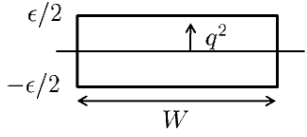}}
\caption{Another cross section: effective diffusion coefficient depends on form of cross section.}
\end{figure}

The physical reason why the diffusion coefficient increases at the curved point  is similar to the discussion of total resistance for the  following electric circuit.
Let us consider the parallel connection of three electric resistances.
When the resistances have the same $R$ value , the total resistance is $R/3$.
However, when the resistances have a dispersion $r$ with a mean $R$ value  ($r<<R$),  such as
$R_1 = R+r, ~ R_2=R, $ and $R_3 = R-r$, the total resistance takes the value
$$R_{tot} = \frac{R}{3} \{1 -\frac{2}{3} (\frac{r}{R})^2\}.$$
The dispersion reduces the total resistance, i.e., increases the diffusion coefficient, as is shown in (\ref{eq:eff_diff}).

\section{Beyond local equilibrium condition}

Starting from the local equilibrium distribution, we discuss the perturbation theory.
The fluctuation field $n(t,s,q^2,q^3)$ is introduced as

\begin{equation}
\phi^{(3)} = \frac{\phi^{(1)}(s,t)}{\sigma} + n(s,q^2,q^3,t). \label{eq:fluctuation}
\end{equation}
The normalization condition is
\begin{equation}
0 = \int n \sqrt{G} d\sigma. \label{eq:n_condition}
\end{equation}

Let us introduce the differential equation for the $n$ field.
From (\ref{eq:diff}), (\ref{eq:new}), and (\ref{eq:fluctuation}), we obtain

\begin{eqnarray}
\frac{\partial n}{\partial t} &=& \frac{1}{\sigma} \{ D \hat{\Delta} - \frac{\partial}{\partial s} D_{eff} \frac{\partial}{\partial s} \} \phi^{(1)} \nonumber\\
&+& \{ D \hat{\Delta} n - \frac{D}{\sigma}\int \sqrt{G} \hat{\Delta} n ~d\sigma \}. \label{eq:n_eq}
\end{eqnarray}

By multiplying equation (\ref{eq:n_eq}) by $\sqrt{G}$ and integrating it by $d\sigma$, we obtain
$$\frac{\partial}{\partial t} \int  n \sqrt{G} d\sigma=0.$$

Thus, for consistency,  condition  (\ref{eq:n_condition}) holds anytime when it is satisfied at the initial time.

We solve equation (\ref{eq:n_eq}) by following the method discussed by Zwanzig \cite{Zwanzig}. 
The formal solution of  (\ref{eq:n_eq}) is given as

\begin{eqnarray}
n &=& e^{D \hat{\Delta} t} n(0) + \frac{D}{\sigma} \int_0^t e^{D \hat{\Delta} (t-t')} \nonumber\\
&& [ \hat{F} \phi^{(1)}(t')  - \int \sqrt{G} \hat{\Delta} n(t') ~d\sigma] dt',\label{eq:n_solution}
\end{eqnarray}
where
\begin{equation}
 \hat{F} \equiv  \hat{\Delta} - \frac{\partial}{\partial s} \frac{D_{eff}}{D} \frac{\partial}{\partial s}.
\end{equation}

Hereafter, we utilize the initial condition $n(0)=0$, which satisfies (\ref{eq:n_condition}). 
We solve equation (\ref{eq:n_solution}) by the iterative method.

$$n = n_0 + n_1 + n_2 +\cdots.$$

The 0th order of the iterative solution is

\begin{equation}
n_0 =  \frac{D}{\sigma} \int_0^t e^{D \hat{\Delta} (t-t')} \hat{F} \phi^{(1)}(t') dt',
\end{equation}
where $\exp [D \hat{\Delta} t] \sim \exp [t/(\epsilon^2/D)]$ changes rapidly in a short time scale $\epsilon^2/D$, while $\phi^{(1)}(t)$ changes slowly.
To achieve our calculation,  we utilize the Markov approximation.
For the fast changing variable $S(t)$ and slowly changing variable $M(t)$,
we have for a large $t$,
\begin{equation}
\int_0^t S(t-t') ~M(t')~ dt' \sim  \int_0^\infty S(\tau) ~d\tau ~ M(t).
\end{equation}
Then, we obtain

\begin{eqnarray}
n_0 &=& \frac{D}{\sigma}~  \int_0^\infty e^{D \hat{\Delta} \tau}d\tau ~ \hat{F} \phi^{(1)}(t) \nonumber\\
&=& -\frac{1}{\sigma}~\frac{1}{\hat{\Delta}} \hat{F} \phi^{(1)}(t). \label{eq:n(1)}
\end{eqnarray}

The explicit form of $n_0$ is shown in appendix 2; however, we consider here  more solutions.
The first iterative solution $n_1$ is given by

\begin{eqnarray}
n_1 &=& - \frac{D}{\sigma}~  \int_0^\infty e^{D \hat{\Delta} \tau}d\tau ~ \int \sqrt{G} \hat{\Delta} n_0(t) ~d\sigma \nonumber\\
&=& \frac{1}{\sigma}~ \frac{1}{\hat{\Delta}}\int \sqrt{G} \hat{\Delta} n_0(t) ~d\sigma \nonumber\\
&=&  -\frac{1}{\sigma^2}~ \frac{1}{\hat{\Delta}}\int \sqrt{G} \hat{F} \phi^{(1)}(t) ~ d\sigma = \xi_1,
\end{eqnarray}
where $\xi_1$ is any function that satisfies the Laplace equation.

\begin{equation}
\hat{\Delta} \xi_1 =0. \label{eq:xi}
\end{equation}

This follows the relation
\begin{equation}
\int \sqrt{G}\hat{F} \phi^{(1)} d\sigma
= \int \sqrt{G}(\hat{\Delta} -  \frac{\partial}{\partial s} \frac{D_{eff}}{D} \frac{\partial}{\partial s}) \phi^{(1)} d\sigma =0. \label{eq:0}
\end{equation}

In the same manner, we obtain
\begin{eqnarray}
n_2 &=& - \frac{D}{\sigma}~  \int_0^\infty e^{D \hat{\Delta} \tau}d\tau ~ \int \sqrt{G} \hat{\Delta} \xi_1(t) d\sigma \nonumber\\
&=& \frac{1}{\sigma}~ \frac{1}{\hat{\Delta}}\int \sqrt{G} \hat{\Delta} \xi_1(t) d\sigma = \xi_2,
\end{eqnarray}
where $\xi_2$ satisfies the same equation as $\xi_1$.
We should note that this fact comes from the Markov approximation.
In this manner, we obtain a solution for $n$.

$$ n = n_0 + \xi_1 + \xi_2 + \cdots. $$

From equation (\ref{eq:n(1)}), the general solution of $n_0$ contains the Laplace field that satisfies the same equation as $\xi$. Therefore, the sum of $\xi$ is included in $n_0$. 
Thus, $n_0$ is the exact solution for $n$.

\begin{equation}
n = n_0.
\end{equation}

Furthermore, from equations (\ref{eq:new}) and (\ref{eq:fluctuation}), we have

\begin{equation}
\frac{\partial \phi^{(1)}}{\partial t} = \frac{\partial}{\partial s} D_{eff} \frac{\partial}{\partial s}  \phi^{(1)} + D \int \sqrt{G} \hat{\Delta} n ~ d\sigma . \label{eq:total}
\end{equation}

On the other hand, (\ref{eq:n(1)}) and  (\ref{eq:0}) give

$$\int \sqrt{G} \hat{\Delta} n ~ d\sigma =0.$$

Thus, we observe no effect from fluctuation $n$ even though $n \neq 0$ (see appendix 2).
Although the particles diffuse in the transverse direction, the total flow along the centerline is not affected.
This shows that the effective equation  (\ref{eq:result}) holds even in the case of a nonlocal equilibrium state under the  Markov approximation, where we utilize the condition: $\epsilon^2/D$ is small time scale compared to our time scale $t$.

\section{Mean Square Displacement}

Our quasi-one-dimensional diffusion equation (\ref{eq:result}) determines the time development of MSD as follows.

From the definition of the expectational value, we have

\begin{equation}
<f(s)> \equiv \frac{\int f(s)~\phi(s,t)~ds }{\int \phi(s,t) ~ds}.
\end{equation}

Then, we obtain

\begin{eqnarray}
&&\frac{\partial}{\partial t} <(\Delta s)^2> 
= \frac{\partial}{\partial t}(<s^2> - <s>^2)\nonumber\\
&&~~~ = 2 <D_{eff}(s)> +2 <(\Delta s) D_{eff}'(s)>,
\end{eqnarray}
where $\Delta s \equiv s-<s>$.\\
We also obtain

\begin{eqnarray}
\frac{\partial^2}{\partial t^2} <(\Delta s)^2> &=& 6 <D_{eff}''(s) D_{eff}(s)> \nonumber\\
&+& 2 <D_{eff}'(s)^2>\nonumber\\
&+& 2 <D_{eff}'''(s) D_{eff}(s) (\Delta s)> \nonumber\\
&+& 2<D_{eff}''(s) D_{eff}'(s) (\Delta s)>.
\end{eqnarray}

When the tube forms a constant curvature configuration, such as a torus or a helix,
we obtain

\begin{eqnarray}
<(\Delta s)^2> ~=~ 4D \frac{1-\sqrt{1-(\kappa\epsilon)^2}}{(\kappa \epsilon)^2}~t, \label{eq:msd}
\end{eqnarray}
where
$$\kappa = \frac{1}{R}$$
for a torus with a ring radius $R$ and 

$$ \kappa = \frac{R \omega^2}{\mu^2 +R^2 \omega^2}$$

for a helix defined by

$$x(u)=R \cos \omega u,~y(u)=R \sin \omega u, ~z(u)=\mu u.$$

Here, $u$ is the length parameter

$$ds = \sqrt{\mu^2 + R^2 \omega^2} ~du.$$

On the other hand, if the curvature $\kappa$ exhibits a position dependence, 
we can show the short time expansion for MSD. 
Let us choose the initial condition

\begin{equation}
\phi(t=0,s) = \delta(s).
\end{equation}

Then, we calculate MSD by short time expansion near $t \sim0$.

\begin{equation}
<(\Delta s)^2> = a_1 t + a_2 t^2 + \cdots, \label{MSD2}
\end{equation}

where

\begin{eqnarray}
a_1 &=& \frac{\partial <(\Delta s)^2>}{\partial t}_{t=0}\nonumber\\
&=& 2 <D_{eff}(s)>_{t=0} +2 <(\Delta s) D_{eff}'(s)>_{t=0}\nonumber\\
&=& 2 D_{eff}(0).
\end{eqnarray}

In the same manner, we obtain

\begin{eqnarray}
a_2 &=& \frac{1}{2} \frac{\partial^2 <(\Delta s)^2>}{\partial t^2}_{t=0}\nonumber\\
&=& 3 D_{eff}''(0) D_{eff}(0) +  D_{eff}'(0)^2.
\end{eqnarray}

Here, we utilize

\begin{equation}
D_{eff}(s) = 2D \frac{1-\sqrt{1-(\kappa(s)\epsilon)^2}}{(\kappa(s) \epsilon)^2}.
\end{equation}

Note that these short time expansion coefficients are given by the curvature and its space derivatives. 
This procedure is similar to that given in a theory of diffusion in higher dimensional Riemannian manifold \cite{Pavel},  although the coefficients are determined by the Riemannian curvature that vanishes in our case.
In this manner, we obtain the curvature-dependent MSD for the quasi-one-dimensional curved system.

\section{Conclusion}
We have discussed the effective diffusion equation in a curved thin tube 
with local and nonlocal equilibrium conditions.

We obtained the curvature-dependent diffusion coefficient in the case of the local equilibrium condition. 
The physical interpretation of curvature dependence is given by analogy to Ohm's law.

Next, we studied the case in which the local equilibrium is broken by perturbation, and we showed that the effective diffusion equation does not change under the fluctuation of diffusion in the transversal direction when we work with the Markov approximation. 

We calculated the mean square displacement (MSD) by using the effective diffusion coefficient that depends on the local curvature.
In the case of thin torus and helix, we obtain a constant curvature and a simple solution (\ref{eq:msd})  for MSD.  In general cases, curvature depends on position and then we cannot show the explicit form of MSD; however, we can calculate the short time expansion of MSD in the form (\ref{MSD2}). 
The coefficients are calculated  using the local curvature and its derivatives.
Thus, we have shown the geometrical aspect of the diffusion in the quasi-one-dimensional system.

\section{Appendix 1: Frenet-Seret equations}

Let us consider a line specified by
\begin{equation}
\vec{x}(s),
\end{equation}
where $s$ is the length of the line.
The unit tangent vector is defined by
\begin{equation}
\vec{e}_1 \equiv \frac{d \vec{x}(s)}{ds}.
\end{equation}

Another (normal) unit vector $\vec{e}_2$ is defined by
\begin{equation}
\frac{d \vec{e}_1(s)}{ds} = \kappa \vec{e}_2, ~~ \mid  \vec{e}_2 \mid =1, \label{eq:kappa}
\end{equation}
where $\kappa$ is called the curvature of this line in $R_3$.
The reason for this is as follows.

At point $\vec{x}(s)$, we draw a circle that tangents this curved line.
From figure 5(a), we easily obtain two relations:
\begin{equation}
\vec{e}_1(s+ds) - \vec{e}_1(s) = \vec{e}_2 ~ d \theta, ~~~ ds = R ~d\theta,
\end{equation}
where $R$ is the radius of this circle.
Then, we obtain equation (\ref{eq:kappa}) by identifying $\kappa = 1/R$.

\begin{figure}
\centerline{\includegraphics[width=8cm]{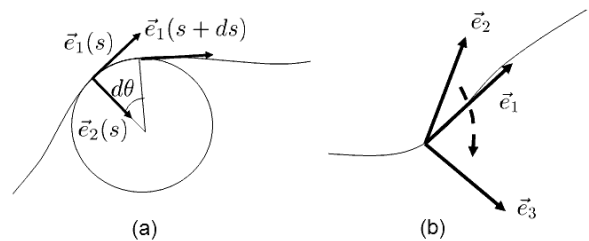}}
\caption{(a) Curvature of line: curved line tangents circle of radius $R$.  (b) Torsion is defined by the rotation of $\vec{e}_2$ around $\vec{e}_1$. }
\end{figure}

Next, we introduce another independent unit vector $\vec{e}_3$ as
\begin{equation}
\vec{e}_3 = \vec{e}_1 \times \vec{e}_2.
\end{equation}

Then, we obtain

\begin{equation}
\frac{d \vec{e}_2}{ds}  = \alpha~ \vec{e}_1 + \tau~ \vec{e}_3, \label{eq:torsion}
\end{equation}
where $\alpha$ and $\tau$ are unknown functions.
However, $\alpha$ can be calculated as

\begin{equation}
 \alpha = \vec{e}_1 \cdot \frac{d \vec{e}_2}{ds}  =  -  \vec{e}_2 \cdot \frac{d \vec{e}_1}{ds} = - \kappa.
\end{equation}

Thus, we replace (\ref{eq:torsion}) with
\begin{equation}
\frac{d \vec{e}_2}{ds}  = - \kappa~ \vec{e}_1 + \tau~ \vec{e}_3.
\end{equation}
The function $\tau$ is called torsion, which has a simple geometrical meaning.

From the equation

\begin{equation}
 \tau = \frac{d \vec{e}_2}{ds}  \cdot  \vec{e}_3,
\end{equation}
we observe that  $\tau$ indicates the rotation of the unit vector $\vec{e}_2$ around the tangential direction of the line, as shown in figure 5(b).

We further consider a derivative of $\vec{e}_3$,

\begin{equation}
\frac{d \vec{e}_3}{ds}  = \beta~ \vec{e}_1 + \gamma~ \vec{e}_2,
\end{equation}
with the unknown functions $\beta$ and $\gamma$. 
However, both functions can be obtained as

\begin{equation}
\beta = \frac{d \vec{e}_3}{ds}  \cdot \vec{e}_1 = -\frac{d \vec{e}_1}{ds}  \cdot \vec{e}_3 = -\kappa \vec{e}_2 \cdot  \vec{e}_3=0,
\end{equation}

and

\begin{equation}
\gamma = \frac{d \vec{e}_3}{ds}  \cdot \vec{e}_2 = -\frac{d \vec{e}_2}{ds}  \cdot \vec{e}_3 = -\tau.
\end{equation}

Thus, we obtain

\begin{equation}
\frac{d \vec{e}_3}{ds}  = - \tau~ \vec{e}_2.
\end{equation}

Let us summarize our results.
We show the following three Frenet-Seret equations and 
two geometrical quantities $\kappa$: curvature and $\tau$: torsion.
\begin{eqnarray}
\frac{d \vec{e}_1}{ds} &=& \kappa \vec{e}_2, \\
\frac{d \vec{e}_2}{ds}  &=& - \kappa~ \vec{e}_1 + \tau~ \vec{e}_3, \\
\frac{d \vec{e}_3}{ds}  &=& - \tau~ \vec{e}_2.
\end{eqnarray}

These three unit orthogonal vectors form the local basis in $R_3$.
Note that $\vec{e}_2$ has the direction of the center of curvature.

\section{Appendix 2: Form of $n$ field}

Let us calculate $n_0$ in the $\epsilon$ expansion under the Markov approximation.
In this section, we utilize simpler notations of local coordinates instead of general coordinates:
$$ q^2 = v = r \cos \theta, ~ q^3 = w = r \sin \theta, ~d\sigma = rdrd\theta = dvdw.$$

\begin{eqnarray}
n_0 = -\frac{1}{\sigma}~\frac{1}{\hat{\Delta}} \hat{F} \phi^{(1)}(t).
\end{eqnarray}

First, we expand $\hat{F}$ in powers of $\epsilon$.
Since $\hat{F}$ acts only on the function of $s$, we obtain
\begin{eqnarray}
 \hat{F} &=& \frac{\partial}{\partial s} \{2\kappa v + 3 \kappa^2 v^2 + 4 \kappa^3 v^3 -(\kappa \epsilon/2)^2  \} \frac{\partial}{\partial s} ~~~~ \nonumber\\
&+& ( \kappa \tau w -\kappa,_{s} v ) (1+ 3\kappa v + 6 \kappa^2 v^2 ) \frac{\partial}{\partial s} + {\cal O}(\epsilon^4)\nonumber\\
&\sim& 2\kappa v \frac{\partial^2}{\partial s^2}  + ( \kappa \tau w +\kappa,_{s} v )\frac{\partial}{\partial s}.
\end{eqnarray}

Then, we observe that $\hat{F}$ starts from ${\cal O}(\epsilon^1)$.
Next, the Laplace-Beltrami operator has the form

\begin{eqnarray}
\hat{\Delta} =&& \hat{\Delta}^{(2)} \nonumber\\
&+& \kappa (\frac{\sin \theta}{r} \frac{\partial}{\partial \theta} - \cos \theta \frac{\partial}{\partial r})
\nonumber\\
&+& \frac{\partial^2}{\partial s^2} -2 \tau \frac{\partial}{\partial s} \frac{\partial}{\partial \theta} + (\kappa^2 \sin \theta \cos \theta - \tau,_{s})\frac{\partial}{\partial \theta} \nonumber\\
&-& \kappa^2 r \cos^2 \theta \frac{\partial}{\partial r} + \tau^2 \frac{\partial^2}{\partial \theta^2} \nonumber\\
&+& {\cal O(\epsilon)},
\end{eqnarray}

where
$$\hat{\Delta}^{(2)} \equiv \frac{\partial^2}{\partial r^2} + \frac{1}{r} \frac{\partial}{\partial r} + \frac{1}{r^2} \frac{\partial^2}{\partial \theta^2} = \frac{\partial^2}{\partial v^2} + \frac{\partial^2}{\partial w^2} $$
is the leading term of ${\cal O}(\epsilon^{-2})$.
Therefore, $\hat{\Delta}^{-1}$ starts from ${\cal O}(\epsilon^{2})$ and then ${\cal O}(\epsilon^{3})$, and so on.
\begin{equation}
\frac{1}{\hat{\Delta}} = \frac{1}{\hat{\Delta}^{(2)}}+ {\cal O}(\epsilon^{3}).
\end{equation}
For simplicity, we calculate $n$ up to ${\cal O}(\epsilon^{1})$.
Then,

\begin{eqnarray}
n_0 &=& -\frac{1}{\sigma}\frac{1}{\hat{\Delta}} \hat{F} \phi^{(1)}\nonumber\\
&\sim& -\frac{1}{\sigma}\frac{1}{\hat{\Delta}^{(2)}}\{2\kappa v \frac{\partial^2}{\partial s^2} + ( \kappa \tau w +\kappa,_{s} v )\frac{\partial}{\partial s}\}\phi^{(1)}.
\end{eqnarray}

Then, we need to calculate the two functions

\begin{eqnarray}
f&=& \frac{1}{\hat{\Delta}^{(2)}} v,\\
g&=& \frac{1}{\hat{\Delta}^{(2)}}w.
\end{eqnarray}
Both functions should satisfy the boundary condition for $n$: $\partial_r n|_{r=\epsilon}=0$.
First, we determine the solutions of 
$$\hat{\Delta}^{(2)} f = v, ~~ \hat{\Delta}^{(2)} g = w.$$
The solutions are
$$f = a v^3 + (1/2 -3a) w^2v + b (v^2-w^2) + c vw + dv+ew +h,$$
$$g = m w^3 +(1/2-3m) v^2w + n(w^2-v^2) + p vw + qw + rv + u,$$
where Latin letters except $v$ and $w$ indicate constants.
The boundary condition specifies the constants
$$f= \frac{v^3}{8} + \frac{vw^2}{8} -\frac{3 \epsilon^2}{8} v +h,$$
$$g= \frac{w^3}{8} + \frac{wv^2}{8} -\frac{3 \epsilon^2}{8} w +u.$$

\begin{figure}
\centerline{\includegraphics[width=9cm]{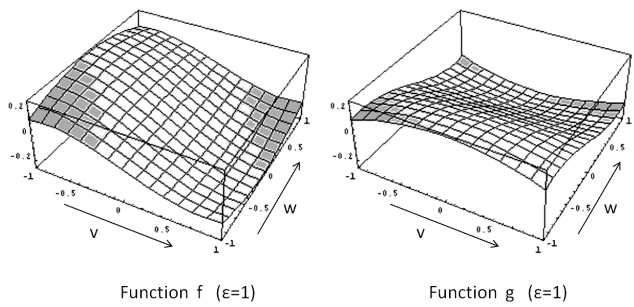}}
\caption{Distributions of $f$ and $g$ with $\epsilon=1$.}
\end{figure}

Then, we obtain
\begin{eqnarray}
n_0&\sim&-\frac{1}{\sigma} \frac{1}{\hat{\Delta}^{(2)}}\{2\kappa v \frac{\partial^2}{\partial s^2}  + ( \kappa \tau w +\kappa,_{s} v ) \frac{\partial}{\partial s}  \}\phi^{(1)} \nonumber\\
&\sim& -\frac{1}{\sigma} [2 \kappa f \frac{\partial^2}{\partial s^2} + \{ \kappa \tau g + \kappa,_{s} f \} \frac{\partial}{\partial s} ]\phi^{(1)}.
\end{eqnarray}
$h$ and $u$ are determined from the normalization condition $\int n \sqrt{G} d\sigma =0$.
Then, we obtain
\begin{equation}
h= -\frac{7}{96}\kappa \epsilon^4, u=0.
\end{equation}

We obtain the final form for $n$ up to ${\cal O}(\epsilon^1)$:
\begin{eqnarray}
n_0
&\sim& -\frac{1}{\sigma} [2 \kappa (\frac{v^3}{8} + \frac{vw^2}{8} -\frac{3 \epsilon^2}{8} v )\frac{\partial^2}{\partial s^2} \nonumber\\
&+& \{ \kappa \tau ( \frac{w^3}{8} + \frac{wv^2}{8} -\frac{3 \epsilon^2}{8} w) \nonumber\\
&+& \kappa,_{s} (\frac{v^3}{8} + \frac{vw^2}{8} -\frac{3 \epsilon^2}{8} v ) \} \frac{\partial}{\partial s}]\phi^{(1)}.
\end{eqnarray}

\begin{acknowledgments}
The author would like to thank the referee of PLA for giving him very important suggestions and  Professor Giga of Tokyo University and Professor Yokoyama of Gakushuin University for encouragement and discussions.
\end{acknowledgments}

\end{document}